\documentclass[]{kluwer}    % Specifies the document style.
\usepackage{graphicx}
\usepackage{klups}

\begin{document}
\input psfig.tex 
\begin{article}

\begin{opening} 	
\title{Homogeneity of early--type galaxies across clusters}
\author{Stefano \surname{Andreon}}  
\runningauthor{Stefano Andreon}
\runningtitle{Homogeneity of early--type galaxies across clusters}
\institute{INAF--Osservatorio Astronomico di Capodimonte \\
INAF--Osservatorio Astronomico di Brera}
\date{10 Dec 2002}

\begin{abstract} 
We studied the scatter across clusters  of the color of the red sequence in a
representative and large sample of clusters (more than 200) detected on the
Early Data Release of the Sloan Digital Sky Survey (EDR--SDSS) in the redshift
range $0.06<z<0.34$. We found an extreme degree of homogeneity in the color of
the red sequence (the intrinsic scatter is about 0.02 to 0.03 mag) suggesting
that either galaxies on the red sequence formed a long time ago ($z>2$) or
else their star formation is universally delayed with preservation of a small
spread in age formation. The latter possibility is ruled out by the mere
existence of galaxies at high redshift. While the old age of ellipticals was
already been claimed for a small heterogeneous collection of clusters, most of
which are rich ones, we found that it holds for a ten to one hundred larger
sample, representative of all clusters and groups detected on the EDR--SDSS.
Hence, we claim the possible universality of the color of the galaxies on the
red sequence. Furthermore, the sample includes a large number of very poor
clusters (also called groups), not studied in previous works, for which the
hierarchical and monolithic scenarios of elliptical formation predict
different colors for the brightest ellipticals. The observed red sequence
color does not depend on cluster/group richness at a level of 0.02 mag, while
a 0.23 mag effect is expected according to the hierarchical prediction.
Therefore, the stellar population of red sequence galaxies is similar in
clusters and groups, in spite of different halo histories. Finally, since the
observed rest--frame color of the red sequence does not depend on environment
and redshift, it can be used as a distance indicator, with an error
$\sigma_z=0.018$, a few times better than the precision achieved by other
photometric redshift estimates and twice better than the precision of the
Fundamental Plane for a single galaxy at the median redshift of the EDR-SDSS. 
\end{abstract}  \end{opening}		

\section{Introduction} 
 
The existence of a relation between color and magnitude for early type galaxies
has been known for a long time (e.g. Sandage \& Visvanathan 1978). 
This relation,
known as color--magnitude relation, implies a link between the mass of the
stellar population and its age or metallicity, with the latter being the
presently favored explication, because high redshift clusters have bluer
color--magnitude relations with a small scatter (Ellis et al. 1997, Kodama \&
Arimoto 1997, Andreon, Davoust \& Heim 1997, Stanford, 
Eisenhardt \& Dickinson 1998).
The color--magnitude relation, often called red sequence, of
different clusters shows an homogeneity across clusters, but from
the observational point of view  
the sample studied thus far is a small heterogeneous collection of
clusters, and not a representative sample of clusters in the Universe.

At first sight, the early formation of the bulk of stars in cluster galaxies 
seems a problem for hierarchical scenarios of galaxy formation. However,
while star formation occurs late in such models, it is pushed back to early
times in clusters (Kauffmann 1996). Hence, the conclusion that  the
bulk of stars in cluster galaxies is homogeneous and red  
is not in contradiction with
the hierarchical scenario. A natural consequence of hierarchical models
would be
that the evolution of early--type galaxies should depend on their environment:
merging history are likely to have varied from cluster to groups.

\section{The data \& the analysis}

We used the EDR--SDSS galaxy catalog (Stoughton et al. 2002).   We detected
clusters on the EDR-SDSS in a way  not biasing the cluster catalog against (or
for) clusters with a normal or unusual color for their galaxies (e.g. with
early--type galaxies having unusually red or blue colors). 
Details are presented
in Andreon (2003). The detected clusters span a large
range in richness: there are many clusters of very low richness (often call
groups), so low that they are not listed in the Abell (1958) catalog (i.e.
$R<0$), and also very rich clusters clusters ($R=3$ or higher).

Of the $\sim450$ clusters detected in the redshift range $0.06<z<0.34$, we are
able to measure their reshift (using the spectroscopic  {\it galaxy} database
of the EDR-SDSS), for about half to two thirds of them. With the most stringent
criteria, used hereafter, 212 clusters have spectroscopic redshift. This
cluster sample is larger than previous studied samples (by a factor of 10 to
100), and is  representative of the whole cluster population detected on the
EDR--SDSS in the $0.06<z<0.34$ range (except for clusters so rare that their
inclusion in our sample is unlikely), because it is a large and random (color
unbiased) subsample of the whole sample.

For each cluster, we measure the color of the red sequence (the intercept of
the color--magnitude relation), by a method that ensures that the color is
measured with a fixed accuracy, independent on the cluster richness (details 
are presented in Andreon et al. 2003). Finally, we
measure the difference between the observed color of the red sequence and the
average color at each redshift.

\begin{figure*}
%\centerline{\includegraphics[width=8truecm]{andreon_fig1.ps}}
\psfig{figure=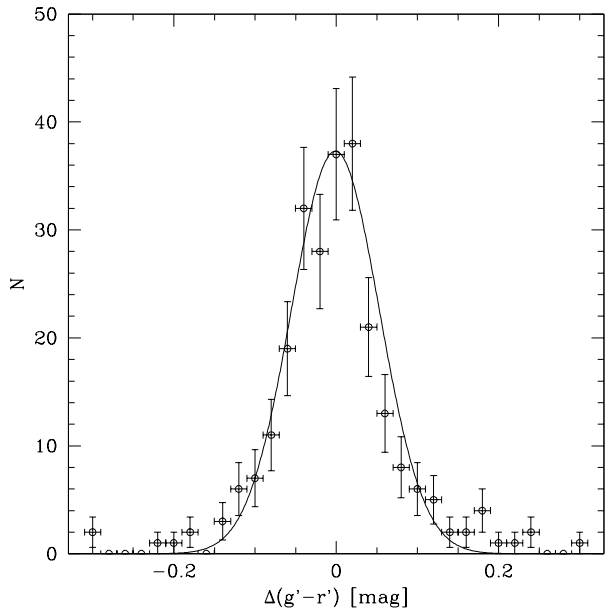,height=8truecm}
\caption[h]{Distribution of the difference between observed and average
color of the red sequnce. The
curve is a Gaussian of the same median and (robust) dispersion of the data,
traced to guide the eye, but never used in the paper.
Error bars on data points are $\sqrt{N}$ (in the ordinate direction) and
half the bin width (in the abscissa direction)}
\end{figure*}

\section{Results}

Figure 1 shows  the difference between observed and average color. The
dispersion is 0.054 mag.  We checked if part of this scatter is due
to systematic effects, such as color drift across the survey. We were unable to
see any trend between color residuals and observing run, CCD id, right
ascension or declination, hence supporting the non--systematic nature of the
scatter.  Part of the observed scatter is of observational nature, not
intrinsic to variations, from cluster to cluster, of the color of the red
sequence.  The intrinsic scatter of the color of the red sequence across
clusters  turns out to be about 0.02-0.03 mag. 

We build a model that describes the observations with the aim of putting
constrains on the synchronicity or stochasticity of the 
stellar formation in galaxies on
the red sequence. We inspired ourselves on Bower, Lucey \& Ellis (1992 and
following works): via a stellar population 
synthesis model we infer the mean age
of the last episode of star--formation and its scatter by requiring that the
expected color dispersion across clusters matched the observed one. We use
the scatter alone, and not the color variation, the former being more 
robust than the
latter to systematic uncertainties of stellar population models. 
The small scatter implies that either galaxies on the red
sequence formed a long time ago ($z>2$) or else their star formation is
universally delayed with preservation of a small spread in age formation. The
latter possibility is ruled out by the mere existence of galaxies at high
redshift (e.g. Shapley et al. 2001).

\begin{figure*}
\centerline{\psfig{figure=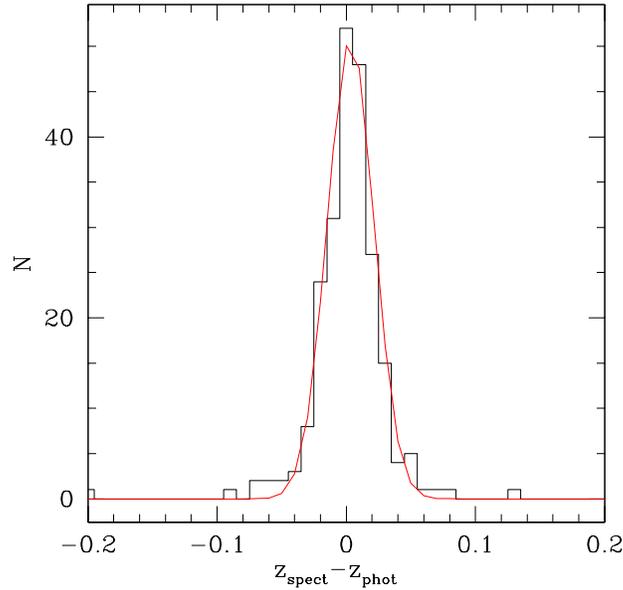,height=8truecm}}
\caption[h]{Residual between photometric and spectroscopic
redshifts. The curve shows a Gaussian with the same width as the observed 
distribution and with the same area. }
\end{figure*}

While the old age of ellipticals had already been claimed for small
heterogeneous collections of rich clusters, we found
that it holds for a ten to one hundred larger sample, representative of all
clusters and groups detected on the EDR--SDSS. Hence, we claim the possible
universality of the color of the galaxies on the red sequence at $0.06<z<0.34$.
Furthermore, the sample includes a large number of very poor clusters (also
called groups), not studied in previous works, for which the hierarchical and
monolithic scenarios of elliptical formation predict different colors for the
brightest ellipticals. 

The observed red sequence color does not depend on cluster/group richness at a
level of 0.02 mag, while a 0.23 mag effect is expected according to the
hierarchical prediction. Therefore, the stellar population of red sequence
galaxies is similar in clusters and groups, in spite of different halo
histories. As their cousins in clusters, galaxies in groups live an
``accelerated" live, in spite of the prediction of the hierarchical scenario.

Since the color of the red sequence does not depend on redshift and
environment, it can be used as a distance indicator. The scatter between
$z_{spect}$ and
$z_{photom}$, determined from the color of the red sequence, 
turns out to be 0.018 in $z$ between $0.06<z_{photom}<0.30$, as shown in
Figure 2. The considered redshift range is a bit smaller than in the remaining
of the paper,  because the flattening of the color--redshift relation at
$z>0.3$ does not allow to accurately measure $z_{photom}$ at $z>0.3$. The
observed scatter (0.018), computed on about 200 clusters with $z_{spect}$,
outperform by a factor of 3 all photometric redshift precisions published thus
far (usually 0.05 or larger) and by a factor two the precision of the
Fundamental Plane for a single galaxy (e.g. Jorgensen, Franx, \& Kjaergaard
1996, 0.03 at the median redshift of our survey).

\acknowledgements 
I acknowledge partial funding from the conveners of the ws-ge
for financial support through project PESO/PRO/15130/1999 
from FCT/Portugal which has allowed
me to present this work at the JENAM2002 meeting.  A special thank goes to the
EDR--SDSS collaboration for their excellent work, and to Alfred P. Sloan for his
gift to the astronomy. The standard acknowledgement for researches based on
EDR--SDSS data can be found in  the EDR--SDSS Web site: http://www.sdss.org/.

\end{article}
\end{document}